\documentclass[12pt]{article}
\usepackage{fancyhdr}
\renewcommand{\maketitle}{
    \begin{center}
      \Large
        {\bf Duality in Off-Shell Electromagnetism}
        \vskip .3 true cm
      \normalsize
        Martin Land \\
        \vskip .3 true cm
        Department of Computer Science \\
        Hadassah College \\
        P. O. Box 1114, Jerusalem 91010, Israel
      \end{center}
      \vskip .5 true cm
}
\input{tcilatex}

\begin{document}

\title{}
\author{}
\maketitle

\begin{abstract}
In this paper, we examine the Dirac monopole in the framework of Off-Shell
Electromagnetism, the five dimensional U(1) gauge theory associated with
Stueckelberg-Schrodinger relativistic quantum theory. After reviewing the
Dirac model in four dimensions, we show that the structure of the five
dimensional theory prevents a natural generalization of the Dirac monopole,
since the theory is not symmetric under duality transformations. It is shown
that the duality symmetry can be restored by generalizing the
electromagnetic field strength to an element of a Clifford algebra.
Nevertheless, the generalized framework does not permit us to recover the
phenomenological (or conventional) absence of magnetic monopoles.
\end{abstract}

\baselineskip7mm \parindent=0cm \parskip=10pt

\section{Introduction}

\subsection{Duality}

Dirac's 1931 model for the magnetic monopole \cite{monopole}, opened a
number of interesting directions for research. Aside from providing a
possible explanation for the quantization of electric charge, Dirac's work
led to investigation of the topological aspects of gauge theory, and
eventually, to the study of duality relations in non-Abelian and
supersymmetric gauge theories. These areas of contemporary research emerged
from Dirac's work because the seemingly simple question, `Why is the
magnetic field sourceless, while the electric field may be induced by a
source?' touches on subtle aspects of the dimensionality of spacetime and
the structure of admissible field theories. Although the question of the
magnetic monopole can be approached by simply adding a magnetic current that
makes Maxwell's equations symmetric under exchange of ${\mathbf{E}}$ and ${%
\mathbf{H}}$, this symmetry relies on the duality relation between the two
tensor equations, 
\begin{eqnarray}
\partial_\nu F^{\mu\nu}&=&J^\mu  \label{0} \\
\epsilon ^{\mu \nu \lambda \rho }\partial _{\nu }F_{\lambda \rho}&=&
\partial _{\nu }\Bigl[\epsilon ^{\mu \nu \lambda \rho }F_{\lambda \rho}%
\Bigr] = \partial _{\nu }\tilde{F}^{\mu \nu } = 0   \label{1}
\end{eqnarray}
and the fact that $\tilde{F}^{\mu \nu }$ is equivalent to $F^{\mu\nu}$ with $%
{\mathbf{E}} \rightarrow -{\mathbf{H}}$ and ${\mathbf{H}}\rightarrow {%
\mathbf{E}}$. The homogeneous Bianchi relation (\ref{1}) connects the
inhomogeneous equation (\ref{0}) to the global gauge invariance associated
with current conservation and to the local gauge invariance that leads to
potential theory. These deep features of the Maxwell theory provide the
machinery that underly the symmetry under exchange of ${\mathbf{E}}$ and ${%
\mathbf{H}}$, and the only `accidental' element of this symmetry is that the
theory is written in four dimensions where the Levi-Civita density has four
indices, and hence the dual of the second rank field strength tensor is also
a second rank tensor.

In this paper, we examine the question of duality in the framework of
Off-Shell Electromagnetism \cite{saad}, the five dimensional U(1) gauge
theory associated with the Stueckelberg-Schrodinger relativistic quantum
theory \cite{Stueckelberg}. Although the five dimensions constitute a vector
and a scalar representation of the four dimensional Lorentz group, the
structure of the theory prevents a natural generalization of the Dirac
monopole based on duality. We first show that the five dimensional
generalization of the Maxwell equations do not possess duality symmetry.
Subsequently, we demonstrate how the symmetry can be restored by
generalizing the framework to a direct sum of tensors of multiple rank in a
Clifford algebra. Finally, we show that although the duality symmetry is
restored, the generalized framework does not permit the magnetic current to
be `rotated away' in order to recover the phenomenological (or conventional)
absence of monopoles. We leave the topological aspects of the theory and the
question of charge quantization to a subsequent paper.

\subsection{Off-Shell Electromagnetism}

Off-Shell Electromagnetism is a five dimensional generalization of the
standard Maxwell theory, to which it reduces in an equilibrium limit. The
field equations are 
\begin{equation}
\partial _{\beta }f^{\alpha \beta }(x,\tau )=ej^{\alpha }(x,\tau )\ \ \ \ \
\ \ \ \ \ \ \ \ \ \ \ \ \ \ \epsilon ^{\alpha \beta \gamma \delta
\varepsilon }\partial _{\alpha }f_{\beta \gamma }(x,\tau )=0  \label{100}
\end{equation}%
where $\alpha ,\beta ,\gamma ,\delta ,\varepsilon =0,1,2,3,5$. The five
indices correspond to the four dimensions of spacetime $x^{\mu }$, where $%
\mu ,\nu ,\lambda =0,1,2,3$, and a Poincar\'{e} invariant parameter $%
x^{5}=\tau $ that labels events along particle worldlines. The
\textquotedblleft fifth dimension\textquotedblright\ is formally similar to
the Galilean invariant time in Newtonian theory, generalized to a covariant
theory in which worldlines are traced out by the classical four-vector $%
x^{\mu }\left( \tau \right) $ or the quantum wave function $\psi (x,\tau )$,
as the parameter proceeds monotonically from $\tau =-\infty $ to $\tau
=\infty $. The fields act on particles through a covariant form of the
classical Lorentz force originally proposed by Stueckelberg \cite{Stueckelberg}, 
\begin{eqnarray}
M\;\ddot{x}_{\mu } &=&\lambda e\ f_{\mu \alpha }(x,\tau )\,\dot{x}^{\alpha
}=\lambda e\ 
\Bigl%
[f_{\mu \nu }(x,\tau )\,\dot{x}^{\nu }+f_{\mu 5}(x,\tau )%
\Bigr%
]  \nonumber \\
\frac{d}{d\tau }\left( -\frac{1}{2}M\dot{x}^{2}\right) &=&\lambda e\
f_{5\alpha }(x,\tau )\,\dot{x}^{\alpha }=\lambda e\ f_{5\mu }(x,\tau )\,\dot{%
x}^{\mu } \ \ .  \label{120}
\end{eqnarray}%
This clasical theory can be derived \cite{emlf} from the quantum theory proposed by Sa'ad,
Horwitz and Arshansky \cite{saad} 
\begin{equation}
\Bigl[i\partial _{\tau }+\lambda ea_{5}(x,\tau )\Bigr]\psi (x,\tau )=\frac{1%
}{2M}\Bigl[p^{\mu }-\lambda ea^{\mu }(x,\tau )\Bigr]\Bigl[p_{\mu }-\lambda
ea_{\mu }(x,\tau )\Bigr]\psi (x,\tau )\;\;,  \label{130}
\end{equation}%
where the potentials are related to the field strengths through 
\begin{equation}
f_{\alpha \beta }=\partial _{\alpha }a_{\beta }-\partial _{\beta }a_{\alpha
}\;.
\end{equation}%
The dimensional constant $\lambda $ is required for consistency with Maxwell
theory, as shown below. Equations (\ref{100}) to (\ref{120}) have been
obtained \cite{beyond} as the most general classical theory consistent with
the commutations relations 
\begin{equation}
\left[ x^{\mu },x^{\nu }\right] =0\qquad m\left[ x^{\mu },\dot{x}^{\nu }%
\right] =-i\hbar g^{\mu \nu }\left( x\right) \;,  \label{140}
\end{equation}%
and as the dynamical theory associated with the invariance of (\ref{130})
under local gauge transformations of the type 
\begin{equation}
\psi (x,\tau )\rightarrow e^{i\lambda e\Lambda (x,\tau )}\psi (x,\tau )%
\mbox{\qquad}\mbox{\qquad}a_{\alpha }(x,\tau )\rightarrow a_{\alpha }(x,\tau
)+\partial _{\alpha }\Lambda (x,\tau )\;\;.
\end{equation}%
The inclusion of the parameter $\tau $ in the gauge function \cite{saad}
generalizes other parameterized quantum theories \cite{Stueckelberg, others}%
, leading to a well-posed and integrable electrodynamics of interacting
events. Relaxing the mass-shell constraint in (\ref{140}) breaks general
reparameterization invariance, but under the conditions 
\begin{equation}
f_{5\mu }=0\qquad \mathrm{and}\qquad \partial _{\tau }f^{\mu \nu }=0\;\;,
\label{170}
\end{equation}%
the remaining $\tau $-translation symmetry is associated, via Noether's
theorem, with dynamic conservation of the mass. It has been shown \cite{emlf}
that while the material events and gauge fields may exchange mass when the
conditions (\ref{170}) do not hold, the total mass-energy of the particles
and fields is conserved. Since the gauge fields propagate with a mass
spectrum, this theory has been called off-shell electrodynamics. Equation (%
\ref{130}) admits the five dimensional conserved current 
\begin{equation}
\partial _{\mu }j^{\mu }+\partial _{\tau }j^{5}=\partial _{\mu }\frac{-i}{2M}%
\Bigl[\psi ^{\ast }(\partial ^{\mu }-i\lambda ea^{\mu })\psi -\psi (\partial
^{\mu }+i\lambda ea^{\mu })\psi ^{\ast }\Bigr]+\partial _{\tau }\Bigl|\psi %
\Bigr|^{2}=0
\end{equation}%
\smallskip leading to the interpretation of $\Bigl|\psi (x,\tau )\Bigr|^{2}$
as the probability density at $\tau $ of finding the event at the spacetime
point $x$. The connection with Maxwell theory is made by extending an
observation by Stueckelberg: if (\ref{170}) holds asymptotically, pointwise
in $x$ as $\tau \rightarrow \pm \infty $, then integration of (\ref{100})
over $\tau $, called concatenation of events into a worldlines \cite{concat}%
, recovers standard Maxwell theory 
\begin{equation}
\partial _{\nu }F^{\mu \nu }=eJ^{\mu }\qquad \qquad \epsilon ^{\mu \nu \rho
\lambda }\partial _{\mu }F_{\nu \rho }=0
\end{equation}%
where 
\begin{equation}
J^{\mu }(x)=\int_{-\infty }^{\infty }d\tau \;j^{\mu }(x,\tau )\mbox{\quad}%
\mathrm{and}\mbox{\quad}F^{\mu \nu }(x)=\int_{-\infty }^{\infty }d\tau
\;f^{\mu \nu }(x,\tau )  \label{200}
\end{equation}%
and so $f^{\mu \nu }(x,\tau )$ has been called the pre-Maxwell field. It
follows from (\ref{200}) that $\lambda $ has dimensions of length.

\subsection{Spacetime Algebra}

The spacetime algebra formalism \cite{Hestenes} achieves a high degree of
notational compactness by representing the usual tensorial objects of
physics as intrinsic (basis independent, and hence index-free) elements in a
Clifford algebra. Among the structural features of Clifford algebra that
enable multiple physical statements to be combined are the direct sums of
tensors of various rank and the natural separation of the product into a
rank-lowering symmetric part and a rank-raising antisymmetric part.
Furthermore, the close connection between Clifford numbers and geometry, has
led to recent work \cite{Matej-et-al} suggesting a generalization of
standard physical entities through their representation as Clifford numbers.
In this paper, we make use of both features of Clifford algebra, and here we
introduce the most basic elements of the formalism.

Following Hestenes, we begin with the space of vectors $a,b,c,...$
consisting of real-valued $D$-tuples 
\begin{equation}
a=\left( a_{0},a_{1},\cdots ,a_{D-1}\right)
\end{equation}%
in Minkowski space with 
\begin{equation}
g=\mathrm{diag}\left( -1,1,\cdots ,1\right) \;\;.
\end{equation}%
To construct the Clifford algebra $\mathcal{C}_{D}$ we introduce a
multiplication operation, with the requirement that Clifford numbers $A,B,C,...$,
consisting of sums of products of any number of vectors, satisfy

\begin{enumerate}
\item $A+B=B+A$

\item $A+\left( B+C\right) =\left( A+B\right) +C$

\item $A\left( BC\right) =\left( AB\right) C$

\item $A\left( B+C\right) =AC+BC$

\item $ab=a\cdot b$, if and only if $a$ and $b$ are collinear.
\end{enumerate}

From these requirements, one may derive \cite{thesis} the rules of the
spacetime algebra $\mathcal{C}_{D}$, a subset of which we give here without
proof. The product of two vectors separates naturally into a symmetric part
and antisymmetric part 
\begin{equation}
ab=\frac{1}{2}\left( {ab+ba}\right) +\frac{1}{2}\left( {ab-ba}\right)
=a\cdot b+a\wedge b
\end{equation}%
where the symmetric part can be identified with the scalar inner product,
and the rank 2 antisymmetric part is called a bivector. The general Clifford
number is a direct sum of multivectors of rank $0,1,\ldots ,D$ 
\begin{eqnarray}
A &=&A_{0}+A_{1}+A_{2}+A_{3}+\cdots +A_{D}  \label{230} \\
&=&A_{0}+A_{1}^{i}{\mathbf{e}}_{i}+A_{2}^{ij}{\mathbf{e}}_{i}\wedge {\mathbf{%
e}}_{j}+\cdots +A_{D}^{i_{0}i_{2}\cdots i_{D-1}}{\mathbf{e}}_{i_{0}}\wedge {%
\mathbf{e}}_{i_{1}}\wedge \cdots \wedge {\mathbf{e}}_{i_{D-1}}
\end{eqnarray}%
expanded on the basis 
\begin{equation}
\left\{ {1,{\mathbf{e}}_{i},{\mathbf{e}}_{i}\wedge {\mathbf{e}}_{j},{\mathbf{%
e}}_{i}\wedge {\mathbf{e}}_{j}\wedge {\mathbf{e}}_{k},\cdots ,{\mathbf{e}}%
_{0}\wedge {\mathbf{e}}_{1}\wedge \cdots \wedge {\mathbf{e}}_{D-1}}\right\}
\;\;.
\end{equation}%
The terms in (\ref{230}) admit a geometric interpretation: $A_{1}$
represents an oriented line, $A_{2}$ represents an oriented plane, $A_{3}$
represents an oriented volume, and so on. From the antisymmetry of the wedge
product it follows that the $r$-product 
\begin{equation}
{{\mathbf{e}}_{0}\wedge {\mathbf{e}}_{1}\wedge \cdots \wedge {\mathbf{e}}_{r}%
}
\end{equation}%
in $D$ dimensions spans a $\binom{D}{r}$ dimensional subalgebra, and the
dimension of the full Clifford algebra is 
\begin{equation}
\dim \mathcal{C}_{D}=\sum\limits_{r=0}^{D}\binom{D}{r}=2^{D}\;\;.
\label{254}
\end{equation}%
The product of a vector and an $r$-vector similarly separates into a
symmetric part and antisymmetric part%
\begin{equation}
aA_{r}=a\left( {a_{1}\wedge a_{2}\wedge \cdots \wedge a_{r}}\right) =a\cdot
A_{r}+a\wedge A_{r}  \label{260}
\end{equation}%
where 
\begin{eqnarray}
a\cdot A_{r} &=&\frac{1}{2}\left[ {aA_{r}-\left( {-1}\right) ^{r}A_{r}a}%
\right]  \\
&=&\sum\limits_{i=1}^{r}{\left( {-1}\right) ^{i+1}\left( {a\cdot a_{i}}%
\right) ~}a_{1}\wedge \cdots \wedge a_{i-1}\wedge a_{i+1}\wedge \cdots
\wedge a_{r}  \label{280} \\
a\wedge A_{r} &=&\frac{1}{2}\left[ {aA_{r}+\left( {-1}\right) ^{r}A_{r}a}%
\right] =a\wedge a_{1}\wedge a_{2}\wedge \cdots \wedge a_{r}\;\;\;\;\;\;\;\;.
\end{eqnarray}%
The geometric interpretation of the separation in (\ref{260}) is seen by
taking a unit vector $a^{2}=1$. With a little algebra, it can be shown that 
\begin{eqnarray}
a\cdot \left[ a\left( a\cdot A{_{r}}\right) \right]  &=&a\cdot A{_{r}=a}%
\cdot A_{r}^{\Vert } \\
a\cdot \left[ a\left( a\wedge A{_{r}}\right) \right]  &=&0={a}\cdot
A_{r}^{\bot }
\end{eqnarray}%
and it follows that the symmetric and antisymmetric parts of the product
correspond to the linearly dependent and orthogonal components of the
multivector $A{_{r}}$ 
\begin{equation}
A{_{r}}=a^{2}A{_{r}}=a\left( a\cdot A{_{r}}\right) +a\left( a\wedge A{_{r}}%
\right) =A_{r}^{\Vert }+A_{r}^{\bot }\;\;.  \label{295}
\end{equation}%
In component form, linear independence of ${\mathbf{e}}_{k}$ and $A{_{r}}$
requires that $A_{r}^{j_{1}\cdots j_{r}}=0$ when any of its indices takes
the value $k$.

It follows from (\ref{254}) that the $D$-vector in $D$ dimensions spans a
one-dimensional subalgebra, and so the element 
\begin{equation}
i={\mathbf{e}}_{0}\wedge {\mathbf{e}}_{1}\wedge \cdots \wedge {\mathbf{e}}%
_{D-1}
\end{equation}%
is known as the unit pseudoscalar, and satisfies 
\begin{equation}
i^{2}=\left( {-1}\right) ^{\frac{{D\left( {D-1}\right) }}{2}}g_{00}\cdots
g_{D-1,D-1}\;\;.  \label{310}
\end{equation}%
The unit pseudoscalar maps an $r$-vector to its dual multivector, a $\left(
D-r\right) $-vector, through 
\begin{equation}
i\left[ {{\mathbf{e}}_{i_{1}}\wedge \cdots \wedge {\mathbf{e}}_{i_{r}}}%
\right] =g_{i_{1}i_{1}}\cdots g_{i_{r}i_{r}}\frac{1}{\left( D-r\right) {!}}{%
\ \varepsilon^{i_{1}\cdots i_{r} \ i_{r+1}\cdots i_{D}}\left[ {\ {\mathbf{e}}%
_{i_{r+1}}\wedge \cdots \wedge {\mathbf{e}}_{i_{D}}}\right] }
\end{equation}%
providing the following useful relationships between the symmetric and
antisymmetric products 
\begin{eqnarray}
a\cdot \left( {iA_{r}}\right) &=&\frac{1}{2}\left[ {aiA_{r}-\left( {-1}%
\right) ^{D-r}}\left( {iA_{r}}\right) {a}\right]  \label{330} \\
&=&\frac{1}{2}\left[ {\left( {-1}\right) ^{D-1}iaA_{r}-\left( {-1}\right)
^{D-r}}\left( {iA_{r}}\right) {a}\right]  \label{340} \\
&=&\left( {-1}\right) ^{D-1}\frac{1}{2}\left[ {iaA_{r}+\left( {-1}\right)
^{r}iA_{r}a}\right] \\
&=&\left( {-1}\right) ^{D-1}i\left( {a\wedge A_{r}}\right)  \label{360}
\end{eqnarray}%
and%
\begin{eqnarray}
a\wedge \left( {iA_{r}}\right) &=&\frac{1}{2}\left[ {aiA_{r}+\left( {-1}%
\right) ^{D-r}}\left( {iA_{r}}\right) {a}\right] \\
&=&\frac{1}{2}\left[ {\left( {-1}\right) ^{D-1}iaA_{r}+\left( {-1}\right)
^{D-r}}\left( {iA_{r}}\right) {a}\right] \\
&=&\left( {-1}\right) ^{D-1}\frac{1}{2}\left[ {iaA_{r}-\left( {-1}\right)
^{r}iA_{r}a}\right] \\
&=&\left( {-1}\right) ^{D-1}i\left( {a\cdot A_{r}}\right) \mbox{\qquad}%
\mbox{\qquad}\mbox{\qquad}\mbox{\qquad}.  \label{368}
\end{eqnarray}%
In writing (\ref{330}) it was necessary to use the known rank of $\left( {%
iA_{r}}\right) $ in fixing the sign ${\left( {-1}\right) ^{D-r}}$, and in
the transition from (\ref{330}) to (\ref{340}) it was necessary to use the
known rank and symmetry of $ia$ in fixing the sign ${\left( {-1}\right)
^{D-1}}$. The appearance of the common factor $\left( {-1}\right) ^{D}$
indicates the role of duality in these simple identities.

In analogy to complex numbers, the general Clifford number (\ref{230}) can
be written%
\begin{eqnarray}
A &=&A_{0}+A_{1}+A_{2}+\cdots +A_{D-2}+A_{D-1}+A_{D} \\
&=&A_{0}+A_{1}+A_{2}+\cdots +\tilde{A}_{2}i+\tilde{A}_{1}i+\tilde{A}_{0}i \\
&=&\left( A_{0}+\tilde{A}_{0}i\right) +\left( A_{1}+\tilde{A}_{1}i\right)
+\left( A_{2}+\tilde{A}_{2}i\right) +\cdots
\end{eqnarray}%
where $\tilde{A}_{0}$ is the scalar dual to $A_D$, $\tilde{A}_{1}$ is the
vector dual to $A_{D-1}$, $\tilde{A}_{2}$ is the bivector dual to $A_{D-2}$,
and so on.

\section{Duality in Maxwell's Equations}

\subsection{Electrodynamics in Spacetime Algebra}

In the spacetime algebra formalism, the Maxwell equations in $D$-dimensions
assume the compact form 
\begin{equation}
dF=J\;\;.  \label{370}
\end{equation}%
As in the standard tensor formulation, the electromagnetic field strength is
a bivector (antisymmetric second rank tensor), whose component form in four
dimensions is%
\begin{eqnarray}
F &=&\dfrac{1}{2}F^{\alpha \beta }\left( \mathbf{e}_{\alpha }\wedge \mathbf{e%
}_{\beta }\right) \\
&=&\left[ F^{01}\left( \mathbf{e}_{0}\wedge \mathbf{e}_{1}\right)
+F^{02}\left( \mathbf{e}_{0}\wedge \mathbf{e}_{2}\right) +F^{03}\left( 
\mathbf{e}_{0}\wedge \mathbf{e}_{3}\right) \right.  \nonumber \\
&&\mbox{\qquad\qquad}\left. +F^{12}\left( \mathbf{e}_{1}\wedge \mathbf{e}%
_{2}\right) +F^{13}\left( \mathbf{e}_{1}\wedge \mathbf{e}_{3}\right)
+F^{23}\left( \mathbf{e}_{2}\wedge \mathbf{e}_{3}\right) \right] \\
&=&\left[ E^{1}\left( \mathbf{e}_{0}\wedge \mathbf{e}_{1}\right)
+E^{2}\left( \mathbf{e}_{0}\wedge \mathbf{e}_{2}\right) +E^{3}\left( \mathbf{%
e}_{0}\wedge \mathbf{e}_{3}\right) \right.  \nonumber \\
&&\mbox{\qquad\qquad}\left. +H^{3}\left( \mathbf{e}_{1}\wedge \mathbf{e}%
_{2}\right) -H^{2}\left( \mathbf{e}_{1}\wedge \mathbf{e}_{3}\right)
+H^{1}\left( \mathbf{e}_{2}\wedge \mathbf{e}_{3}\right) \right] \ \ \ \ \ .
\label{369}
\end{eqnarray}%
The gradient and current are written as vectors $d=\partial ^{\alpha }%
\mathbf{e}_{\alpha }$ and $J=J^{\alpha }\mathbf{e}_{\alpha }$. Equation (\ref%
{370}) splits into the symmetric and antisymmetric parts 
\begin{equation}
d\cdot F+d\wedge F=J
\end{equation}%
which can be separated by rank into 
\begin{eqnarray}
&&d\cdot F=J  \label{430} \\
&&d\wedge F=0\;\;\;.  \label{440}
\end{eqnarray}%
From (\ref{280}) it follows that the components of (\ref{430}) are the
tensor form of the inhomogeneous Maxwell equations, which can be seen from%
\begin{equation}
d\cdot F=\partial ^{\alpha }\left( \dfrac{1}{2}F^{\beta \gamma }\right) 
\mathbf{e}_{\alpha }\cdot \left( \mathbf{e}_{\beta }\wedge \mathbf{e}%
_{\gamma }\right) =\partial ^{\alpha }F^{\beta \gamma }\dfrac{1}{2}\left(
g_{\alpha \beta }\mathbf{e}_{\gamma }-g_{\alpha \gamma }\mathbf{e}_{\beta
}\right) =\left( \partial _{\beta }F^{\beta \gamma }\right) \mathbf{e}%
_{\gamma }\;\;.
\end{equation}%
The total antisymmetry of $\mathbf{e}_{\alpha }\wedge \mathbf{e}_{\beta
}\wedge \mathbf{e}_{\gamma }$ shows that (\ref{440}) is the homogeneous
Maxwell equation.

The dual to $F$ is the multivector $\tilde{F}=iF$, given by%
\begin{equation}
iF=\dfrac{1}{2}F^{\alpha \beta }\left( \mathbf{e}_{0}\mathbf{e}_{1}...%
\mathbf{e}_{D-1}\right) \left( \mathbf{e}_{\alpha }\wedge \mathbf{e}_{\beta
}\right) =\dfrac{1}{2}\dfrac{1}{\left( D-2\right) !}~\epsilon ^{\beta \gamma
\alpha _{1}\alpha _{2}...\alpha _{D-2}}F_{\beta \gamma }\mathbf{e}_{\alpha
_{1}}\mathbf{e}_{\alpha _{2}}...\mathbf{\ e}_{\alpha _{D-2}}
\end{equation}%
so that in four dimensions, the dual is also a bivector 
\begin{eqnarray}
iF &=&\dfrac{1}{4}\epsilon ^{\alpha _{1}\alpha _{2}\beta \gamma }F_{\beta
\gamma }\mathbf{e}_{\alpha _{1}}\wedge \mathbf{e}_{\alpha _{2}} \\
&=&\left[ \epsilon ^{0123}F_{01}\mathbf{e}_{2}\wedge \mathbf{e}_{3}+\epsilon
^{0231}F_{02}\mathbf{e}_{3}\wedge \mathbf{e}_{1}+\epsilon ^{0312}F_{03}%
\mathbf{e}_{1}\wedge \mathbf{e}_{2}\right.  \nonumber \\
&&\mbox{\qquad\qquad}\left. +\epsilon ^{1230}F_{12}\mathbf{e}_{3}\wedge 
\mathbf{e}_{0}+\epsilon ^{2310}F_{23}\mathbf{e}_{1}\wedge \mathbf{e}%
_{0}+\epsilon ^{3120}F_{31}\mathbf{e}_{2}\wedge \mathbf{e}_{0}\right] \\
&=&\left[ E_{1}\mathbf{e}_{2}\wedge \mathbf{e}_{3}-E_{2}\mathbf{e}_{1}\wedge 
\mathbf{e}_{3}+E_{3}\mathbf{e}_{1}\wedge \mathbf{e}_{2}\right.  \nonumber \\
&&\mbox{\qquad\qquad}\left. -H_{3}\mathbf{e}_{0}\wedge \mathbf{e}_{3}-H_{1}%
\mathbf{e}_{0}\wedge \mathbf{e}_{1}-H_{2}\mathbf{e}_{0}\wedge \mathbf{e}_{2}%
\right] \ \ \ \ \ .  \label{485}
\end{eqnarray}%
Equation (\ref{485}) is recognized as the field strength tensor (\ref{369})
with the electric and magnetic fields exchanged. Using the relation (\ref%
{360}), equation (\ref{440}) can also be written 
\begin{equation}
i\left( d\wedge F\right) =d\cdot \left( iF\right) =d\cdot \tilde{F}=0\;\;.
\label{460}
\end{equation}

\subsection{Duality of the Electric and Magnetic Fields in $D=4$}

Following Dirac's famous argument, we may enhance the symmetry of Maxwell's
equations by adding to (\ref{460}) a non-zero current vector. Then,
Maxwell's equations become%
\begin{eqnarray}
d\cdot F &=&J_{electric}  \label{462} \\
d\cdot \tilde{F} &=&J_{magnetic}  \label{464}
\end{eqnarray}%
and in light of (\ref{460}), (\ref{464}) may be rewritten as a trivector
relation%
\begin{eqnarray}
d\cdot \tilde{F} &=&d\cdot \left( iF\right) =i\left( d\wedge F\right)
=J_{magnetic} \\
i^{2}\left( d\wedge F\right)  &=&{-}\left( d\wedge F\right) =iJ_{magnetic} \\
d\wedge F &=&-iJ_{magnetic}\;\;.  \label{472}
\end{eqnarray}%
Combining (\ref{462}) and (\ref{472}) we find%
\begin{equation}
d\cdot F+d\wedge F=dF=J_{electric}-iJ_{magnetic}=J\;\;,  \label{474}
\end{equation}%
in which the current Clifford number $J$ is now a direct sum of a vector and
trivector. Equations (\ref{462}) and (\ref{464}) are form invariant under
the exchange 
\begin{eqnarray}
F &\leftrightarrow &\tilde{F} \\
J_{electric} &\leftrightarrow &J_{magnetic}
\end{eqnarray}%
and equivalently (\ref{474}) is invariant under continuous duality
transformations $U\left( \theta \right) =e^{\theta i}$, which by (\ref{310})
can be written in $D=4$ as 
\begin{equation}
U\left( \theta \right) =e^{\theta i}=\cos \theta +i\sin \theta \;\;.
\end{equation}%
Applied to the field equations (\ref{474})%
\begin{eqnarray}
&&UdF=UJ \\
&&(1+\theta i+o(\theta ^{2}))(d\cdot F+d\wedge F)=(1+\theta i+o(\theta
^{2}))(J_{electric}-iJ_{magnetic})\;\;.
\end{eqnarray}%
Separating terms of equal rank, 
\begin{eqnarray}
&&d\cdot F+\theta id\wedge F=J_{electric}-\theta
i^{2}J_{magnetic}=J_{electric}+\theta J_{magnetic}=J_{electric}^{\prime } \\
&&d\wedge F+\theta id\cdot F=-i\left( J_{magnetic}-\theta
J_{electric}\right) =-iJ_{magnetic}^{\prime }
\end{eqnarray}%
and using relations (\ref{360}) and (\ref{368}), we find form invariance in
the form%
\begin{eqnarray}
&&d\cdot F-\theta d\cdot \left( iF\right) =d\cdot \left( F-\theta \tilde{F}%
\right) =d\cdot F^{\prime }=J_{electric}^{\prime } \\
&&d\wedge F-\theta d\wedge \left( iF\right) =d\wedge \left( F-\theta \tilde{F%
}\right) =d\wedge F^{\prime }=-iJ_{magnetic}^{\prime }\;\;.
\end{eqnarray}%
Dirac argued that the apparent absence of the magnetic current is actually a
convention, according to which we view the world under the specific duality
rotation which takes%
\begin{equation}
J_{magnetic}^{\prime }=\left( J_{magnetic}\right) \cos \theta -\left(
J_{electric}\right) \sin \theta \rightarrow 0\;\;.
\end{equation}%
This argument requires (or implies) that the ratio of electric to magnetic
charge $e/g$ be universal. If the magnetic current exists, then it should
contribute to the classical Lorentz force as%
\begin{equation}
\mathbf{F}=e\left[ \mathbf{E}+\mathbf{v}\times {\mathbf{H}}\right] +g\left[ {%
{\mathbf{H}}-{\mathbf{v}}\times {\mathbf{E}}}\right] \;\;.
\end{equation}%
Writing this Lorentz force in the covariant form%
\begin{equation}
M\frac{{d^{2}x^{i}}}{{d\tau ^{2}}}=eF^{i\mu }\dot{x}_{\mu }+g\epsilon ^{i\mu
\nu \lambda }F_{\mu \nu }\dot{x}_{\lambda }=eF^{i\mu }\dot{x}_{\mu }+g\tilde{%
F}^{i\mu }\dot{x}_{\mu }  \label{492}
\end{equation}%
we see that under the duality rotation, this force becomes 
\begin{equation}
M\frac{{d^{2}x^{i}}}{{d\tau ^{2}}}=e^{\prime }\left( {F^{\prime }}\right)
^{i\mu }\dot{x}_{\mu }
\end{equation}%
consistent with Dirac's argument that the absence of magnetic monopoles is a
convention.

\subsection{Duality of the Electric and Magnetic Fields in $D=5$}

The difficulties in extending Dirac's argument to five dimensions can be
seen in a number of ways, perhaps most simply by observing that the $10$
components of the electromagnetic field strength tensor do not divide
naturally into a pair of vector fields ${\mathbf{E}}$ and ${\mathbf{H}}$ as
occurs in $D=4$. Similarly, the extension of the Lorentz force to five
dimensions, in the form of (\ref{492}), requires that the Levi-Cevita symbol
be extended to five indices. A duality transformation leaving the expression%
\begin{equation}
M\frac{{d^{2}x^{i}}}{{d\tau ^{2}}}=eF^{i\alpha }\dot{x}_{\alpha }+g\tilde{F}%
^{i\alpha }\dot{x}_{\alpha }=eF^{i\alpha }\dot{x}_{\alpha }+g\epsilon
^{i\alpha \beta \gamma \delta }F_{\alpha \beta \gamma }\dot{x}_{\delta }
\label{495}
\end{equation}%
form invariant requires, at minimum the existence of a trivector field
strength $F_{\alpha \beta \gamma }$. In this section, we demonstrate that,
even in the spacetime algebra formalism which allows the representation of a
physical object by a direct sum of various multivectors, Dirac's model for
the monopole in U(1) electrodynamics cannot be reproduced when $D\neq 4$.

Summarizing the basic ingredients of Dirac's model, we require:

\begin{enumerate}
\item Maxwell's equations in the form $dF=J=J_{\left( 1\right) }+J_{\left(
3\right) }$, where the subscript labels the rank of the multivector,

\item A continuous transformation of the form 
\begin{equation}
U=e^{\theta G}=1+\theta G+o(\theta ^{2})
\end{equation}%
generated by some Clifford number $G$,

\item An algebraic structure in which the action of the transformation on
Maxwell's equations%
\begin{equation}
(1+\theta G)(d\cdot F+d\wedge F)=(1+\theta G)(J_{\left( 1\right) }+J_{\left(
3\right) })
\end{equation}%
leads to separation of terms by rank, 
\begin{eqnarray}
&&d\cdot F+\theta Gd\wedge F=J_{\left( 1\right) }+\theta GJ_{\left( 3\right)
} \\
&&d\wedge F+\theta Gd\cdot F=J_{\left( 3\right) }+\theta GJ_{\left( 1\right)
}
\end{eqnarray}

\item A generator $G$ which permits the replacements 
\begin{eqnarray}
&&G\left( d\wedge F\right) =d\cdot \left( GF\right)  \label{550} \\
&&G\left( d\cdot F\right) =d\wedge \left( GF\right)  \label{560}
\end{eqnarray}%
so that the transformed multivectors can be recombined as 
\begin{eqnarray}
d\cdot F+\theta Gd\wedge F &=&d\cdot \left( F+\theta GF\right) =J_{\left(
1\right) }+\theta GJ_{\left( 3\right) } \\
d\wedge F+\theta G\left( d\cdot F\right) &=&d\wedge \left( F+\theta
GF\right) =J_{\left( 3\right) }+\theta GJ_{\left( 1\right) }
\end{eqnarray}

\item Form covariance expressed through the requirement that the pairs
$\{F,GF\}$ and \linebreak $\{J_{\left( 3\right) },GJ_{\left(1\right) }\}$ be of equal
rank.
\end{enumerate}

The problem in five dimensions begin in requirement (4), with finding an
element $G$ of $\mathcal{C}_{N}$ which satisfies (\ref{550}) and (\ref{560}%
). Any choice of $G$, other than a scalar or pseudoscalar, determines an
orthogonal subspace of vectors $\{a_{k}\left\vert \; a_{k}\cdot G=0\right.
\} $, and it follows from (\ref{295}) that the product $GF$ splits into
terms of various rank, according to 
\[
GF=GF^{\Vert }+GF^{\bot } 
\]%
where $a_{k}\cdot F^{\bot }=0$. The definite rank of the product ${iA_{r}}$
was critical to the derivation of (\ref{360}) and (\ref{368}), and
expressions (\ref{550}) and (\ref{560}) can only be satisfied when $G=i$.
This choice implies that requirement (5) can only be satisfied in $D=4$,
where both $F$ and $iF$ are bivectors.

\subsection{Generalized Field}

It was seen in (\ref{495}) that Dirac's model cannot be simply extended to
five dimensions, because the duality transformation takes the bivector field
into a trivector field. We may ask whether the generalization of the
electromagnetic field to include elements of higher rank can restore the
symmetry under duality. In this section, we demonstrate that requirements
(1) to (5) can be satisfied by writing the electromagnetic field tensor as a
general Clifford number. We propose to represent the electromagnetic field
and the source current as a direct sum of multivectors of all possible rank
in $D=5$,%
\begin{eqnarray}
F &=&F_{0}+F_{1}+F_{2}+F_{3}+F_{4}+F_{5} \\
J &=&J_{0}+J_{1}+J_{2}+J_{3}+J_{4}+J_{5}\;\;\;\;\;.
\end{eqnarray}%
The left hand side of Maxwell's equations%
\begin{equation}
dF=J
\end{equation}%
becomes%
\begin{eqnarray}
dF &=&dF_{0}+dF_{1}+dF_{2}+dF_{3}+dF_{4}+dF_{5} \\
&=&dF_{0}+d\cdot F_{1}+d\wedge F_{1}+d\cdot F_{2}+d\wedge F_{2}  \nonumber \\
&&\mbox{\qquad}+d\cdot F_{3}+d\wedge F_{3}+d\cdot F_{4}+d\wedge F_{4}+d\cdot
F_{5}
\end{eqnarray}%
where we used $d\wedge F_{D}\equiv 0$. Equating terms of equal rank, we find
a system of five coupled equations,%
\begin{eqnarray}
d\cdot F_{1} &=&J_{0}  \label{590a} \\
dF_{0}+d\cdot F_{2} &=&J_{1}  \label{590b} \\
d\wedge F_{1}+d\cdot F_{3} &=&J_{2}  \label{590c} \\
d\wedge F_{2}+d\cdot F_{4} &=&J_{3}  \label{590d} \\
d\wedge F_{3}+d\cdot F_{5} &=&J_{4}  \label{590e} \\
d\wedge F_{4} &=&J_{5}\ \ \ .  \label{590f}
\end{eqnarray}%
Writing the duality transformation $U\left( \theta \right) =e^{\theta i}$
for small $\theta $ allows us to collect terms by rank on the right hand
side,%
\begin{equation}
J_{r}\rightarrow J_{r}+\theta iJ_{5-r}\ ,\ \ r=0,\ldots ,5
\end{equation}%
and on the left hand side,%
\begin{eqnarray}
d\cdot F_{1} &\rightarrow &d\cdot F_{1}+\theta id\wedge F_{4} \\
dF_{0}+d\cdot F_{2} &\rightarrow &dF_{0}+d\cdot F_{2}+\theta id\wedge
F_{3}+\theta id\cdot F_{5} \\
d\wedge F_{1}+d\cdot F_{3} &\rightarrow &d\wedge F_{1}+d\cdot F_{3}+\theta
id\wedge F_{2}+\theta id\cdot F_{4} \\
d\wedge F_{2}+d\cdot F_{4} &\rightarrow &d\wedge F_{2}+d\cdot F_{4}+\theta
id\wedge F_{1}+\theta id\cdot F_{3} \\
d\wedge F_{3}+d\cdot F_{5} &\rightarrow &d\wedge F_{3}+d\cdot F_{5}+\theta
idF_{0}+\theta id\cdot F_{2} \\
d\wedge F_{4} &\rightarrow &d\wedge F_{4}+\theta id\cdot F_{1}\ \ \ .
\end{eqnarray}%
Applying (\ref{360}) and (\ref{368}), and collecting terms leads to the
transformed equations 
\begin{eqnarray}
d\cdot \left( F_{1}+\theta iF_{4}\right)  &=&J_{0}+\theta iJ_{5}  \label{600}
\\
d\left( F_{0}+\theta iF_{5}\right) +d\cdot \left( F_{2}+\theta iF_{3}\right)
&=&J_{1}+\theta iJ_{4} \\
d\wedge \left( F_{1}+\theta iF_{4}\right) +d\cdot \left( F_{3}+\theta
iF_{2}\right)  &=&J_{2}+\theta iJ_{3} \\
d\wedge \left( F_{2}+\theta iF_{3}\right) +d\cdot \left( F_{4}+\theta
iF_{1}\right)  &=&J_{3}+\theta iJ_{2} \\
d\wedge \left( F_{3}+\theta iF_{2}\right) +d\cdot \left( F_{5}+\theta
iF_{0}\right)  &=&J_{4}+\theta iJ_{1} \\
d\wedge \left( F_{4}+\theta iF_{1}\right)  &=&J_{5}+\theta iJ_{0}\ \ \ \ \ .
\label{610}
\end{eqnarray}%
Following Dirac, we may identify the transformed equations (\ref{600}) to (%
\ref{610}) with the Maxwell equations in some dual system%
\begin{eqnarray}
d\cdot F_{1}^{\prime } &=&J_{0}^{\prime }  \label{620} \\
dF_{0}^{\prime }+d\cdot F_{2}^{\prime } &=&J_{1}^{\prime } \\
d\wedge F_{1}^{\prime }+d\cdot F_{3}^{\prime } &=&J_{2}^{\prime } \\
d\wedge F_{2}^{\prime }+d\cdot F_{4}^{\prime } &=&J_{3}^{\prime } \\
d\wedge F_{3}^{\prime }+d\cdot F_{5}^{\prime } &=&J_{4}^{\prime } \\
d\wedge F_{4}^{\prime } &=&J_{5}^{\prime }\ \ \ ,  \label{630}
\end{eqnarray}%
and comparing (\ref{620}) --- (\ref{630}) with (\ref{590a}) --- (\ref{590f}%
), it is clear that the duality symmetry has been restored, and that this
procedure can be repeated in any number of dimensions.

However, we cannot claim that in general, the duality transformation that
eliminates the magnetic current $J_{3}^{\prime }$ will eliminate all of the
non-vector currents. To see this, we rewrite (\ref{590c}), (\ref{590e}) and (%
\ref{590f}) in a dual form, by defining%
\begin{eqnarray}
H_{r} &=&iF_{5-r}\ ,\ \ r=1,3,4,5 \\
Q_{s} &=&iJ_{5-s}\ ,\ \ s=2,4,5
\end{eqnarray}%
so that 
\begin{eqnarray}
i\left[ d\wedge F_{1}+d\cdot F_{3}=J_{2}\right] &\rightarrow &d\cdot \left(
iF_{1}\right) +d\wedge \left( iF_{3}\right) =\left( iJ_{2}\right) \\
i\left[ d\wedge F_{3}+d\cdot F_{5}=J_{4}\right] &\rightarrow &d\cdot \left(
iF_{3}\right) +d\left( iF_{5}\right) =\left( iJ_{4}\right) \\
i\left[ d\wedge F_{4}=J_{5}\right] &\rightarrow &d\cdot \left( iF_{4}\right)
=\left( iJ_{5}\right) \ \ \ .
\end{eqnarray}%
become%
\begin{eqnarray}
d\cdot H_{4}+d\wedge H_{2} &=&Q_{3}  \label{635} \\
d\cdot H_{2}+dH_{0} &=&Q_{1}  \label{636} \\
d\cdot H_{1} &=&Q_{0}\ \ \ .  \label{637}
\end{eqnarray}%
Replacing (\ref{590c}), (\ref{590e}) and (\ref{590f}) with (\ref{635}), (\ref%
{636}) and (\ref{637}), the six field equations can be written as%
\begin{eqnarray}
d\cdot F_{1} &=&J_{0} \\
dF_{0}+d\cdot F_{2} &=&J_{1} \\
d\wedge F_{2}+d\cdot F_{4} &=&J_{3} \\
d\cdot H_{1} &=&Q_{0} \\
dH_{0}+d\cdot H_{2} &=&Q_{1} \\
d\wedge H_{2}+d\cdot H_{4} &=&Q_{3}\ \ \ ,
\end{eqnarray}%
so that we are led to two inequivalent sets of coupled field equations of
rank 0, 1, and 3. There is no guarantee that the angle $\theta $ that
transforms $J_{3}\rightarrow J_{3}^{\prime }=0$ will also take $%
Q_{3}\rightarrow Q_{3}^{\prime }=0$. Thus, extending the electromagnetic
field to a general Clifford number does permit the restoration of duality
symmetry, but does not permit the elimination of the magnetic current.

\section{Conclusion}

We have seen that in five dimensions, the usual treatment of electromagnetic
duality cannot be constructed. The duality symmetry which exists in four
dimensions can be minimally restored by generalizing the electromagnetic
field to the Clifford algebra form%
\begin{eqnarray}
F &=&F_{2}+F_{3} \\
J &=&J_{1}+J_{2}+J_{3}+J_{4}\;\;\;\;\;,
\end{eqnarray}%
in which Maxwell's equations%
\begin{equation}
dF=J
\end{equation}%
split into 
\begin{eqnarray}
&&%
\begin{array}{ccr}
d\cdot F_{2}=J_{1} & \mbox{\qquad\qquad} & \ \ \ \ \left( \mathrm{vector}%
\right) 
\end{array}
\label{700} \\
&&%
\begin{array}{ccr}
d\wedge F_{2}=J_{3} & \mbox{\qquad\qquad} & \ \left( \mathrm{trivector}%
\right) 
\end{array}
\label{710} \\
&&%
\begin{array}{ccr}
d\cdot F_{3}=J_{2} & \mbox{\qquad\qquad} & \ \ \ \left( \mathrm{bivector}%
\right) 
\end{array}
\label{720} \\
&&%
\begin{array}{ccr}
d\wedge F_{3}=J_{4} & \mbox{\qquad\qquad} & \left( \mathrm{4-vector}\right) 
\end{array}
\label{730}
\end{eqnarray}%
and the duality transformation $U\left( \theta \right) =e^{\theta i}$
balances (\ref{700}) with (\ref{730}) and (\ref{710}) with (\ref{720})
through%
\begin{eqnarray}
id\cdot F_{2} &=&iJ_{1}\rightarrow d\wedge \left( iF_{2}\right) =\left(
iJ_{1}\right) \mbox{\qquad\qquad}\left( \mathrm{4-vector}\right)  \\
id\wedge F_{2} &=&iJ_{3}\rightarrow d\cdot \left( iF_{2}\right) =\left(
iJ_{3}\right) \mbox{\qquad\qquad}\ \ \ \left( \mathrm{bivector}\right)  \\
id\cdot F_{3} &=&iJ_{2}\rightarrow d\wedge \left( iF_{3}\right) =\left(
iJ_{2}\right) \mbox{\qquad\qquad}\ \left( \mathrm{trivector}\right)  \\
id\wedge F_{3} &=&iJ_{4}\rightarrow d\cdot \left( iF_{3}\right) =\left(
iJ_{4}\right) \mbox{\qquad\qquad}\ \ \ \ \left( \mathrm{vector}\right) \ \ \
\ \ \ .
\end{eqnarray}%
However, this same duality symmetry permits, through the definitions%
\begin{eqnarray}
H_{2} &=&iF_{3} \\
Q_{1} &=&iJ_{4} \\
Q_{3} &=&iJ_{2}\ \ \ ,
\end{eqnarray}%
the identification of (\ref{720}) and (\ref{730}) as a second set of Maxwell
equations%
\begin{eqnarray}
d\cdot H_{2} &=&Q_{1} \\
d\wedge H_{2} &=&Q_{3}\ \ \ \ \ \ ,
\end{eqnarray}%
and in general, the choice of $\theta $ which eliminates $J_{3}$ will not
also eliminate $Q_{3}$. Thus, while the duality symmetry has been restored,
it cannot be exploited to explain the absence of magnetic monopole currents.

The application of Dirac's argument to off-shell electromagnetism requires
that the duality transformation not include the fifth component. The
transformation remains, as in four dimensions, ${\mathbf{E}}\rightarrow {%
\mathbf{H}}$ and ${\mathbf{H}}\rightarrow -{\mathbf{E}}$. In the spacetime
algebra formalism, this transformation is expressed using%
\begin{equation}
i_{4}=\mathbf{e}_{0}\mathbf{e}_{1}\mathbf{e}_{2}\mathbf{e}_{3}
\end{equation}%
which is the unit pseudoscalar in $D=4,$ and in five dimensions is
orthogonal to the unit vector in the 5-direction $\mathbf{e}_{5}$. Since the
choice of orientation which distinguishes the 5-direction is natural to the
off-shell theory, covariance is preserved. It follows from (\ref{485}) that
the transformation which swaps the electric and magnetic fields is%
\begin{eqnarray}
F^{\prime } &=&i_{4}\left( F^{\bot }\right) +F^{\Vert } \\
&=&i_{4}\left[ \mathbf{e}_{5}\left( \mathbf{e}_{5}\wedge F\right) \right] +%
\left[ \mathbf{e}_{5}\left( \mathbf{e}_{5}\cdot F\right) \right]  \\
&=&i_{4}\dfrac{1}{2}F^{\mu \nu }\left( \mathbf{e}_{\mu }\wedge \mathbf{e}%
_{\nu }\right) +F^{\mu 5}\left( \mathbf{e}_{\mu }\wedge \mathbf{e}%
_{5}\right) 
\end{eqnarray}%
where $\mu ,\nu =0,\ldots ,3$. It is interesting to note that since this
transformation leaves the 5-direction unchanged, Dirac's argument for the
absence of magnetic monopole currents requires that $J_{magnetic}^{5}\equiv 0
$, \emph{a priori}, so that it is distinguished from $J_{electric}^{5}$ by the
form of current conservation%
\begin{eqnarray}
\partial _{\alpha }J_{electric}^{\alpha } &=&\partial _{\tau
}J_{electric}^{5}+\partial _{\mu }J_{electric}^{\mu }=0,\ \ \alpha =0,\ldots
3,5 \\
\partial _{\mu }J_{magnetic}^{\mu } &=&0,\ \ \ \ \ \ \ \ \ \ \ \ \ \ \ \ \ \
\ \ \ \ \ \ \ \ \ \ \ \ \ \ \ \ \ \mu =0,\ldots ,3\ \ \ \ \ .
\end{eqnarray}%
While $J_{electric}^{\alpha }$ is understood, in analogy to the classical
non-relativistic case, to be generated by a $\tau $-dependent event density $%
J_{electric}^{5}$ conserved through its flow $J_{electric}^{\mu }$ into
spacetime, the magnetic current appears to be pure spacetime current
circulation, not associated with any endpoint event density. In this sense,
the magnetic current is again distinguished from the electric current by
topological considerations, which we consider in a subsequent paper.

\end{document}